\documentclass[11pt]{article}
\usepackage{graphicx,epsfig}

\newcommand{\BABARPubYear}    {00}

\newcommand{\BABARProcNumber} {31}
\newcommand{\SLACPubNumber} {8742}

\input pubboard/babarsym
\def\result {\ensuremath{\stwob=0.12\pm 0.37\, {\rm (stat)} \pm 0.09\, {\rm (syst)}}}

\def\Title#1{\begin{center} {\Large\bf #1 } \end{center}}
\def\Author#1{\begin{center}{ \large\sc #1} \end{center}}
\def\Address#1{\begin{center}{ \it #1} \end{center}}
\newenvironment{Presented}{\begin{quotation} \begin{center} 
             Presented at the\end{center}
      \begin{center}\begin{large}}{\end{large}\end{center} \end{quotation}}

\begin{document}

% title page
%{\pagestyle{empty}

\begin{flushright}
SLAC-PUB-\SLACPubNumber \\
\babar-PROC-\BABARPubYear/\BABARProcNumber \\
%hep-ex/\LANLNumber \\
January, 2001 \\
\end{flushright}

\def\thefootnote{\fnsymbol{footnote}}

\vfill
\Title{Physics at \Lbabar}

\vfill
\Author{Christos Touramanis\footnote{
Work supported by the U.K. Particle Physics and Astronomy Research Council, Advanced Research Fellowship GR/L04177 }}
\Address{
Department of Physics \\
Oliver Lodge Laboratory, University of Liverpool\\
L69 7ZE , Liverpool , U.K. \\
(for the \lbabar\ Collaboration)
}

\vfill
The \babar\ detector at the SLAC \pep2\ asymmetric \epem\ collider 
has first started collecting data in May 1999. 
A study of time-dependent \CP-violating asymmetries in  
$\Bz \to \jpsi \KS$ and $\Bz \to \psitwos \KS$ decays has been 
performed on a data sample of 9.0\invfb\ taken 
at the \FourS\ resonance and 0.8\invfb\ off-resonance, 
collected through July 2000.  
The preliminary result \result\ is presented here, together with preliminary results 
on neutral and charged \B\ meson lifetimes and \BzBzb\ mixing.

\vfill
\begin{Presented}
5th International Symposium on Radiative Corrections \\ 
(RADCOR--2000) \\[4pt]
Carmel CA, USA, 11--15 September, 2000
\end{Presented}

\vfill
\pagebreak

%%%%%%%%%%%%%%%%%%%%%%   Main document starts here
%
\section{Introduction}
The three--generation Standard Model can accommodate \CP\ violation 
through the presence of a non-zero imaginary phase in the 
Cabibbo-Kobayashi-Maskawa (CKM) quark mixing matrix. 
However, existing measurements of \CP\ violation in the neutral kaon system 
cannot prove that the CKM phase is indeed the origin of \CP\ violation in nature.

The primary goal of the \babar\ experiment at \pep2\ is to elucidate this question 
by a series of observations and measurements of \CP--violating effects in the 
\B\ meson system. These measurements allow the extraction of the angles 
$\alpha$, $\beta$ and $\gamma$ of the Unitarity Triangle, whose non--zero 
area~\cite{Jarlskog} is a direct measure of \CP\ violation.  

\babar\ can also access the sides of the Unitarity Triangle through measurements 
of \Vub, \Vcb in semileptonic \B\ decays and \Vtd\ in \BzBzb\ mixing. This allows 
to overconstrain the Unitarity Triangle and perform stringent tests of 
the Standard Model.

Thus, high statistics, a clean environment and broad access to the rich phenomenology 
of the \B\ sector will allow \babar\ to improve our knowledge of the overall \B\
decay picture and probe New Physics at higher energy scales. A broad heavy flavor 
physics programme is also ongoing in \babar.

%%%%%%%%%%%%%%%%%%%%%%%%%%%%%%%%%%%%%%%%%%%%%%%%%%%%%%%%%%
\section{PEP-II}
The \pep2\ $B$ Factory~\cite{BabarPub0018} is an \epem\ colliding beam storage ring 
complex at SLAC designed to produce a luminosity of 
3x$10^{33} \cm^{-2}s^{-1}$ at a center--of--mass energy of 10.58\gev (\FourS\ resonance). 
During the 2000 run \pep2\ has exceeded this luminosity, while \babar, with a 
logging efficiency of $>$95\%, has routinely accumulated data above its 
design daily rate of $135 \invpb$.

The machine has asymmetric energy beams, with a High Energy Ring 
(HER, 9.0\gev\ electrons) and a Low Energy Ring 
(LER, 3.1\gev\ positrons). These correspond to a 
center--of--mass boost of $\rm {\beta\gamma}$=0.56 and lead to 
an average separation of $\rm {\beta\gamma c \tau}$=250\mum\ 
between the two \B\ mesons vertices, allowing the measurement of 
time--dependent \CP--violating decay rate asymmetries. 

At the \FourS\ resonance \B\ mesons can only be produced as 
{\ensuremath{B^+ {\kern -0.16em B^-}} or coherent 
\BzBzb\ pairs. The time evolution of a coherent \BzBzb\ pair is coupled 
in such a way that the \CP\ or flavor of one \B\ at decay time $t_1$ 
can be described as a function of the other \B\ ($B_{tag}$) flavor at its 
decay time $t_2$ and the signed time difference $\deltat = t_1 - t_2$.

%%%%%%%%%%%%%%%%%%%%%%%%%%%%%%%%%%%%%%%%%%%%%%%%%%%%%%%%%
\section{\babar}
\subsection{Detector description~\cite{BabarPub0018}}
The volume within the 1.5T \babar\ superconducting solenoid contains 
a five layer silicon strip vertex detector (SVT), a central drift
chamber (DCH), a quartz-bar Cherenkov radiation detector (DIRC) and a CsI(Tl) 
crystal electromagnetic calorimeter (EMC). Two layers of
cylindrical resistive plate counters (RPCs) are located between the barrel
calorimeter and the magnet cryostat. 
The instrumented flux return (IFR) outside the cryostat is
composed of 18 layers of radially increasing thickness steel, 
instrumented with 19 layers of planar RPCs in the barrel and 18 in the endcaps 
which provide muon and neutral hadron identification.

%%%%%%%%%%%%%%%%%%%%%%%%%%%%%%%%%%%%%%%%%%%%%%%%%%%%%%%%%%%%
\subsection{Particle reconstruction and identification~\cite{BabarPub0018}}
Charged particle tracking using the SVT and DCH achieves a resolution of 
$\left( \delta p_T / p_T \right)^2 =  (0.0015\, p_T)^2 + (0.005)^2$, 
where $p_T$ is the transverse momentum in \gevc.  
The SVT with a typical resolution of 10\mum\ per hit provides excellent 
vertex resolution  both in the transverse plane and in $z$.  
The typical fully reconstructed single \B\ decay vertex resolution in $z$ is 50\mum. 
Photons are reconstructed in the EMC, yielding mass resolutions  
of 6.9\mevcc\ for \piz\ra\gaga\ and 10\mevcc\ for \KS\ra\piz\piz.

Leptons and hadrons are identified using a combination of measurements
from all the \babar\ components, including 
the energy loss ${\rm d}E/{\rm d}x$ in the helium-based 
gas of the DCH (40 samples maximum) and in the silicon of the SVT (5 samples
maximum). Electron identification is mainly based on the characteristics of their 
shower in the EMC, while muons are identified in the IFR and confirmed by their 
minimum ionising signal in the EMC.
Excellent kaon identification in the barrel region is provided by the DIRC, 
which achieves a separation of $>$3.4$\sigma$ in the range 0.25--3.5\gevc.

%%%%%%%%%%%%%%%%%%%%%%%%%%%%%%%%%%%%%%%%%%%%%%%%%%%%%%%%%%%
\section{\B\ reconstruction}
A variety of inclusive, semiexclusive and exclusive reconstruction methods are applied 
on the \babar\ data, covering semileptonic and pure hadronic decay modes.
The corresponding \B\ samples have different sizes and purity levels and are used for 
different types of studies (Branching Fraction measurements, studies of the dynamics of 
certain decay chains). We will focus here on the cases where some information 
(final state(s), charge, \CP\ or flavor content, decay vertex) can be reconstructed 
for both \B\ mesons in the event.

%%%%%%%%%%%%%%%%%%%%%%%%%%%%%%%%%%%%%%%%%%%%%%%%%%%%%%%%%%%%
\subsection{Exclusive \B\ sample}
\label{sec:exclusive}
\Bz and \Bpm mesons are reconstructed in the following hadronic modes of definite flavor:
\Bz $\to D^{(*)-} \pi^+, D^{(*)-} \rho^+, D^{(*)-} a_1^+$,
\jpsi \Kstarz, \Bub $\to$ \Dz \pim~and \Bub $\to$ \Dstarz \pim\footnote{Throught this paper, 
conjugate modes are implied.}.
All final state particles are reconstructed.
The selections have been optimised for signal significance, using on--peak, off--peak and 
simulated data.
Charged particle identification, mass(or mass difference) and vertex constraints 
are used wherever applicable. 
The signal for each decay mode is identified in the two-dimensional distribution of the
kinematical variables $\Delta E$ and \mes: 
$\Delta E=E^*_{\rm rec}-E^*_b$
is the difference between the \B\ candidate energy and the beam
energy and $\mes=\sqrt{E^{*2}_b - \mbox{\boldmath $p$}^{*2}_{\rm rec}}$ is the
mass of a particle with a reconstructed momentum 
$\mbox{\boldmath $p$}^*_{\rm rec} = \sum_i \mbox{\boldmath $p$}^*_i$
assumed to have the beam energy, as is the case for a true \B\ meson. 
In events with several \B\ candidates only the one
with the smallest $\Delta E$ is considered.
The $\Delta E$ and \mes\ variables have minimal correlation. The
resolution in \mes\ is $\approx$3\mevcc\ and is dominated by 
the beam energy spread. The resolution in
$\Delta E$ is mode dependent 
and varies in the range of 12--40\mev. 
For each mode a rectangular signal region is defined by the three
standard deviation bands in \mes\ ($5.27 < \mes < 5.29$\gevcc)
and $\Delta E$ (mode dependent interval).  
For each mode the sample composition is determined by fitting the
\mes\ distribution for candidates within the
signal region in $\Delta E$ 
to the sum of a single Gaussian representing the signal
and a background function introduced by the ARGUS
collaboration~\cite{bib:argusfunction}. The purity of each subsample
is computed as the ratio of the area of the Gaussian in the $\pm 3
\sigma$ range over the total area in this range. 
Figure~\ref{fig:hadronicb0bch}
shows the \mes\ distributions for the summed  hadronic
\Bz\ and \Bpm\ modes with the fits superimposed.
\begin{figure}[tbhp]
\begin{center}
\begin{tabular}{lr}
\mbox{\epsfxsize=8cm\epsffile{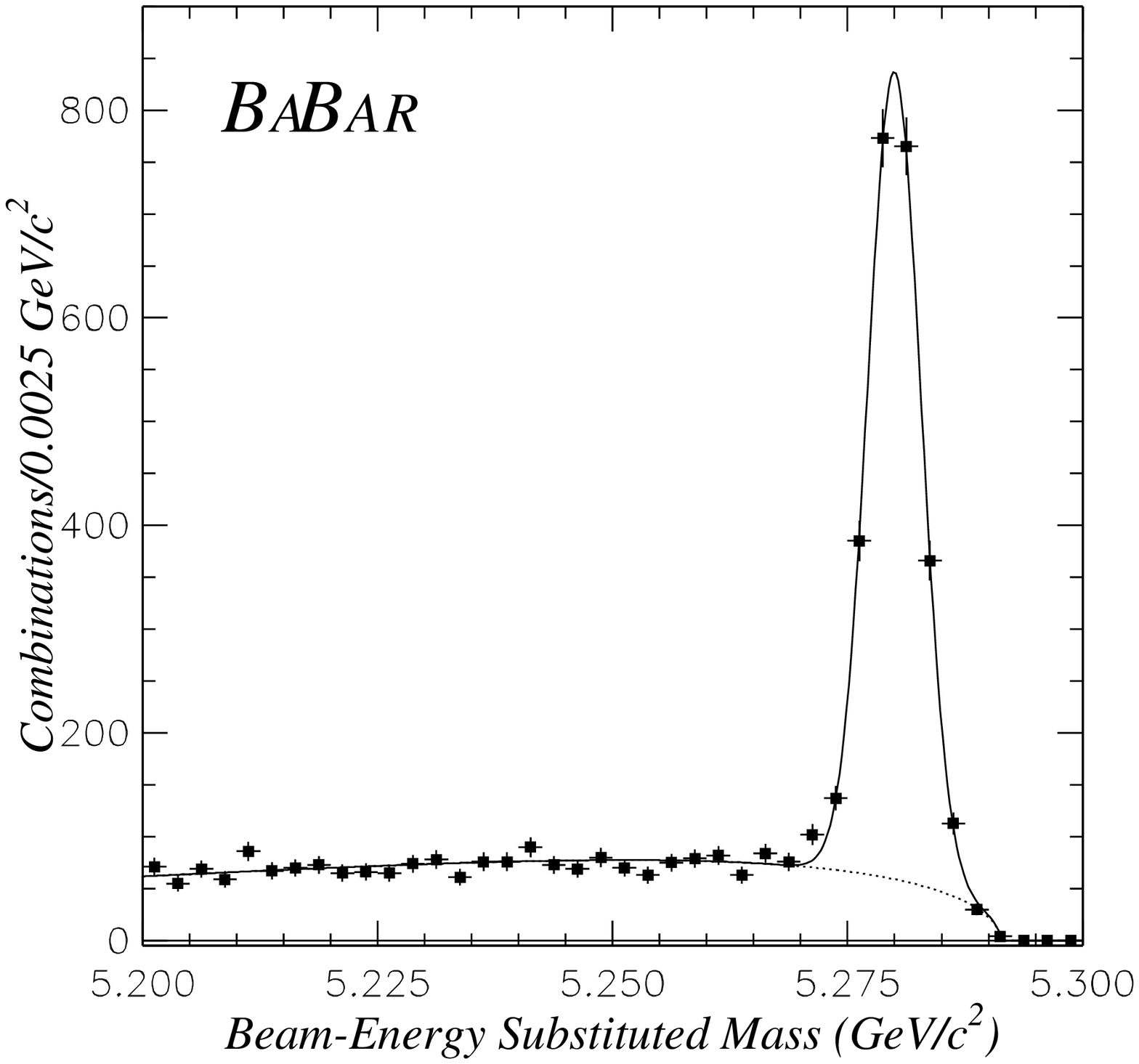}} &
\mbox{\epsfxsize=8cm\epsffile{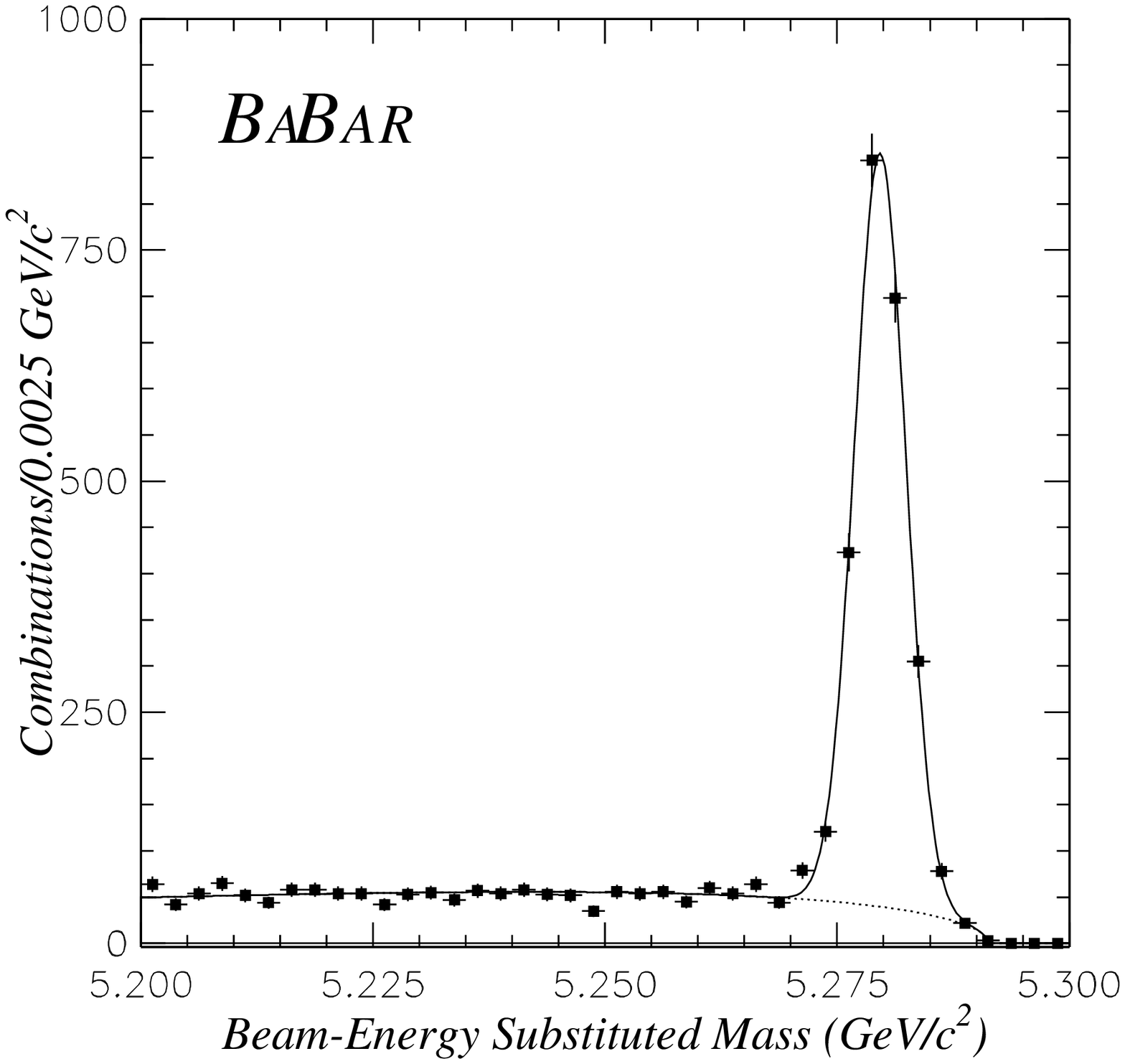}} \\
\end{tabular}
\vskip-6.7cm~~~~~~~~~~~~~~~~~~~~~~~~~~~~~~~~~~~~~~~~~~~~~{\Large
  (a)}~~~~~~~~~~~~~~~~~~~~~~~~~~~~~~~~~~~~~~~~~~~~~~~~~~~~~~~~~~~~~{\Large
  (b)}\\
\vskip6.0cm 
\end{center}
\caption{\mes\ distribution for all the hadronic modes for (a) \Bz\ and 
(b) \Bpm. The complete fit and its ARGUS~\cite{bib:argusfunction} 
background content are also shown. The number of signal events in all 
\Bz\ and \Bpm\ modes are $2577\pm59$ and $2636\pm56$, 
with purity of $\approx$86\% and $\approx$89\% respectively.
\label{fig:hadronicb0bch}}
\end{figure}

%%%%%%%%%%%%%%%%%%%%%%%%%%%%%%%%%%%%%%%%%%%%%%%%%%%%%%%%%
\subsection{Flavor tagging}
\label{sec:tagging}
After removal of the daughter tracks of the reconstructed $B_{rec}$ in an event, 
the remaining tracks are used to determine the flavor of the other \B\ meson 
($B_{tag}$), and this ensemble is assigned a tag flavor, either $\Bz$ or $\Bzb$.  

For each of the tagging methods used we define an effective tagging efficiency  
$Q_i = \varepsilon_i \times \left( 1 - 2\mistag_i \right)^2$, where $\varepsilon_i$ 
is the fraction of events tagged by this method $i$ and 
$\mistag_i$ is the mistag fraction, {\em i.e.} the probability of incorrectly assigning 
the opposite tag to an event using this method. A dilution factor is defined as 
\dilution$= 1 - 2\mistag$ and is extracted from the data for each method.

The {\tt Lepton} category uses the presence and charge of a primary lepton 
from the decaying $b$ quark. Both electrons and muons are used, with 
a minimum center-of-mass momentum requirement of 1.1\gevc. 
If both an electron and a muon candidate satisfy this requirement, only the 
electron is taken into account.
Mistag arises from (a) pions seen as leptons and 
(b) softer opposite-sign leptons coming from charm semileptonic decays.

The {\tt Kaon} category is based on the total charge of all identified Kaons. 
Events with conflicting {\tt Lepton} and {\tt Kaon} tags are excluded from 
both categories.

For events not tagged with the previous methods, a variety of available particle 
identification and kinematic variables are fed in a Neural Network whose 
design and training aims at exploiting the information present in this 
set of correlated quantities. It is sensitive to the presence
of primary and cascade leptons, charged kaons and soft pions from $D^*$ decays. 
In addition, the charge of high-momentum particles is exploited in a 
``jet-charge'' type approach. This functionality has been assigned to different 
sub--nets, to facilitate understanding of the network performance. 
The output from the full neural network tagger $x_{NT}$ is mapped onto the interval 
$\left[ -1, 1 \right]$.  
The assigned flavor tag is \Bz\ if $x_{NT}$ is negative, and \Bzb\ otherwise.
Events with $\left| x_{NT} \right| > 0.5$ are assigned to  the {\tt NT1} tagging category 
and events with  $0.2 < \left| x_{NT} \right| < 0.5$ to the {\tt NT2} tagging category.
Events with  $\left| x_{NT} \right| < 0.2$  have very little tagging 
power and are rejected.

%%%%%%%%%%%%%%%%%%%%%%%%%%%%%%%%%%%%%%%%%%%%%%%%%%%%%%%%%%%
\subsection{\deltat\ calculation and resolution}
\label{sec:deltat}
Since no stable charged particle emerges from the \FourS\ decay point, 
the production point of the \B\ mesons and thus their individual decay 
times cannot be determined. However the decay time difference \deltat\ 
between the two is sufficient for the description of a coherent \B\ meson pair 
(decay length difference technique). 

The event topology is sketched in Fig.~\ref{fig:eventopo}. In the {\em boost 
approximation} used in \babar\ the decay time difference is calculated as : 
$\deltat = \deltaz / c < \beta\gamma >$, where the small flight path of the \B\ 
mesons perpendicular to the z exis is ignored.

Actually, the small effects arising from 
the tilt of the \pep2\ beams with respect to the \babar\ z axis (20\mrad), 
fluctuations in the beam energies, 
the \B\ meson transverse momentum in the \FourS\ rest frame, 
have been studied and are taken into account either in the calculations or in the 
systematic errors as appropriate. 
\begin{figure}[htbp]
\begin{center}
\mbox{\psfig{file=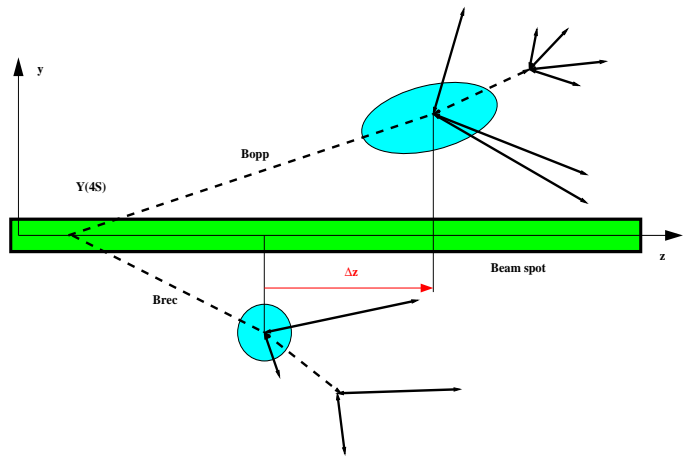,height=6cm,width=9cm}}
\vspace{-0.8cm}
\end{center}
\caption{Event topology showing the two \B\ production and decay points. The figure is not drawn to scale; it has been expanded in the y direction.}
\label{fig:eventopo}
\end{figure}

The resolution \sz\ for the fully reconstructed \B\ is found in the simulation to be 
45--65\mum, depending on the mode. 
The resolution \sz\ for the tag side is $\approx$125\mum, with a small bias of 25\mum\ 
due to forward--going charm decays that cannot be resolved.
The resulting resolution in \deltat\ has been parametrised as the sum of three gaussians. 
The core has a $\sigma$ of 0.6\ps\ and contains 75\% of the events. The tail has a 
$\sigma$ of 1.8\ps. Outliers are described by a gaussian with fixed $\sigma$ of 8\ps, 
that contains $\approx$1\% of the total events. This resolution model is used for the lifetime, 
mixing and \stwob\ fits. Two scale factors (multiplicative to the width of the core and 
tail gaussians) are included in the fits to the real data for the first two cases, 
to account for eventual imperfections in the modeling of $D$ decays and multiple scattering 
in the simulation. 
Extensive studies on the different event samples and with variations of the fits 
(free and fixed parameters) have been performed in order to optimise and validate the method 
and to obtain reliable estimates of the systematic errors. 

%%%%%%%%%%%%%%%%%%%%%%%%%%%%%%%%%%%%%%%%%%%%%%%%%%%%%%%%%%%%%%%%%%%
\section{\B\ lifetime measurements}
\label{sec:lifetime}
The observed \deltat\ distribution for a set of \B\ pair events in the presence of the 
resolution function ${\cal {R}}$ is :
\begin{equation}\label{eq:lifetime}
{\cal F}(\Delta t) = \Gamma \exp(-\Gamma|\Delta t|) \otimes 
{\cal {R}}( \, \deltat \, ; \, \hat {a} \, )
\end{equation}
where $\hat {a}$ is the set of parameters describing the resolution function.

The \B\ meson lifetimes are extracted with unbinned maximum likelihood fits that take 
individual event \deltat\ errors into account. Our preliminary results are :
\begin{eqnarray*}
\tau_{\Bz} &=& 1.506\pm 0.052\ {\rm (stat)} \pm 0.029\ {\rm (syst)}\ \ps \\
\tau_{\Bu} &=& 1.602\pm 0.049\ {\rm (stat)} \pm 0.035\ {\rm (syst)}\ \ps \\
\tau_{\Bu }/\tau_{\Bz } &=& 1.065\pm 0.044 \ {\rm (stat)} \pm 0.021 \ {\rm(syst)}
\end{eqnarray*}
The only background source is combinatorial and it is estimated from the 
side-bands of the beam energy substituted mass variable. The main systematic error 
comes from the resolution modeling and parameters. 
The two proper time fits are shown in Figure~\ref{fig:taub_exclusive}.
\begin{figure}[htbp] 
\begin{center}
  \mbox{\epsfig{file=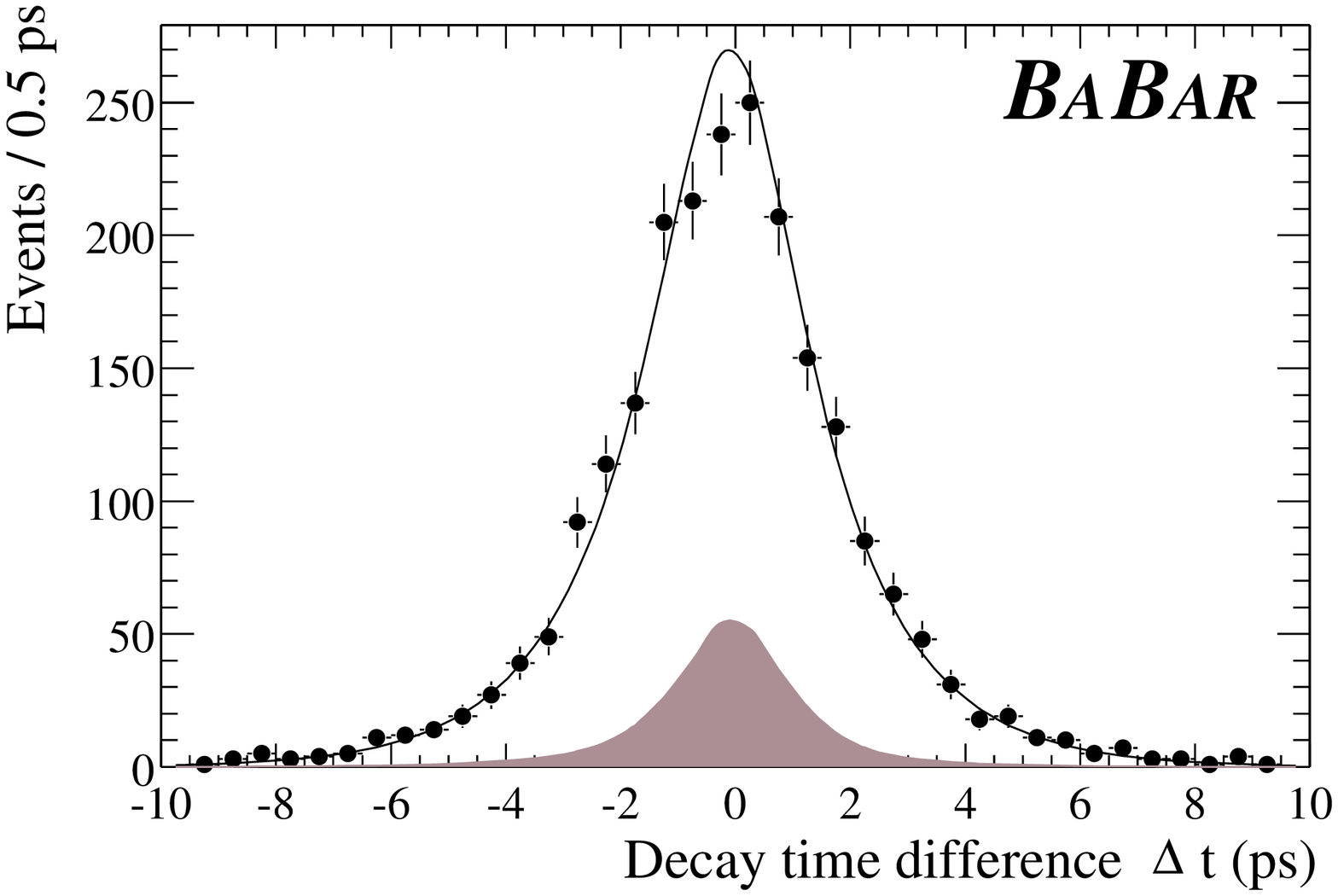,width=8.0cm}}
  \mbox{\epsfig{file=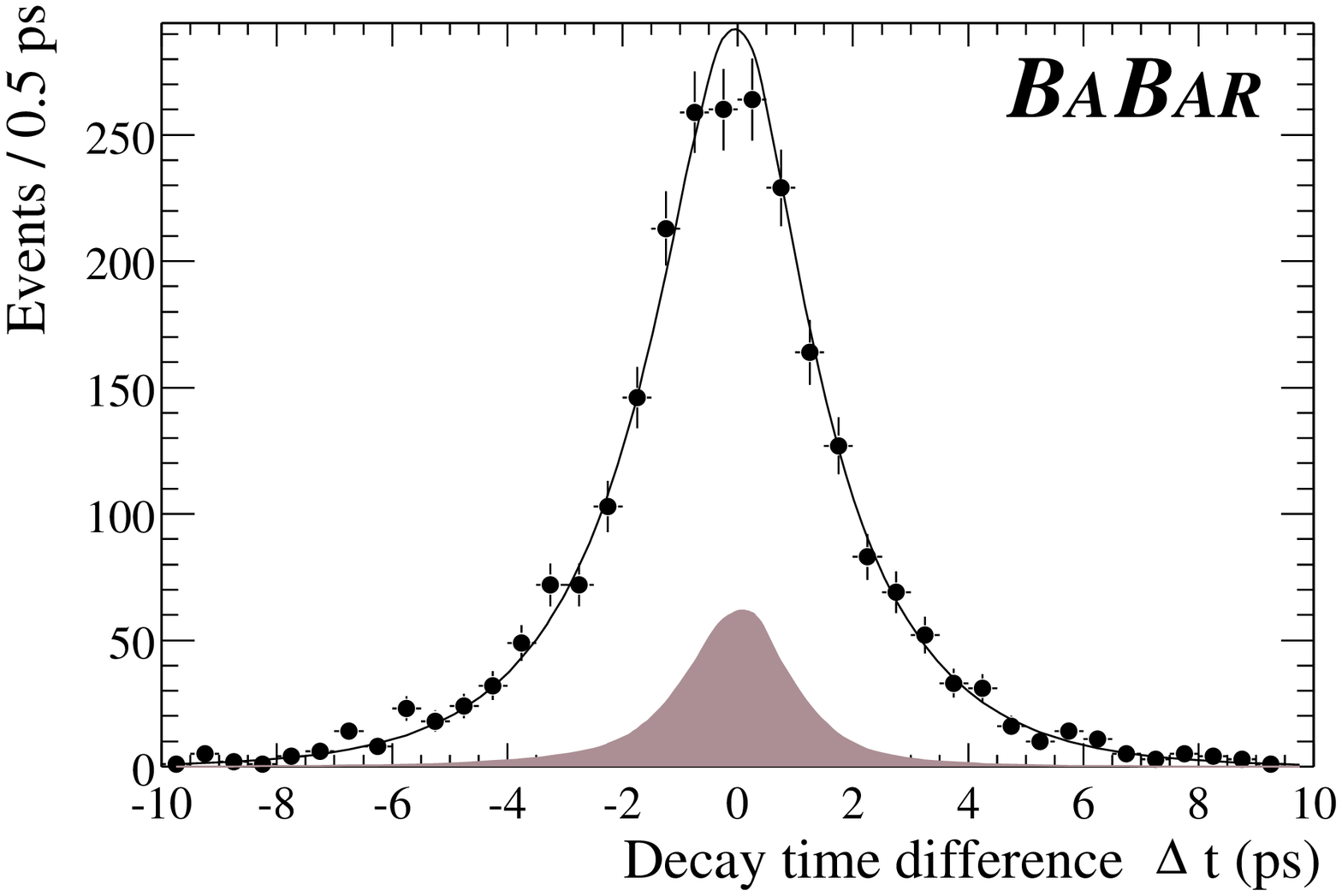,width=8.0cm}}
\vspace{-0.5cm}
\end{center}
\caption{\deltat\ distributions for the \Bz\ (right) and \Bpm\ (left) candidates. 
The result of the lifetime fit is superimposed. 
The hatched areas represent the background content of the event samples.}
\label{fig:taub_exclusive}
\end{figure}

%%%%%%%%%%%%%%%%%%%%%%%%%%%%%%%%%%%%%%%%%%%%%%%%%%%%%%%%%%%%%%%%%%%
\section{\Bz\ mixing measurements}
\label{sec:mixing}
Mixing allows the two neutral \B\ mesons in the \BzBzb\ coherent state 
to decay with the same flavor ({\em mixed events}) 
or the opposite flavor ({\em unmixed events}). In a perfect detector 
one would then observe a time dependent oscillation in the rates of unmixed(+) and 
mixed(-) events :
\begin{equation}
f_\pm( \deltat; \, \Gamma, \, \deltamd ) = 
{\frac{1}{4}}\, \Gamma \, {\rm e}^{ - \Gamma \left| \deltat \right| }
\, \left[  \, 1 \, \pm \, \cos{ \deltamd \, \deltat } \,  \right]
\end{equation}
where \deltamd\ is the difference between the mass eigenstates $B^0_H$ and $B^0_L$.
Due to imperfect tagging and vertex determination the observed rates become :
\begin{equation}
{\cal F}_\pm( \deltat; \, \Gamma, \, \deltamd, \, {\cal {D}}, \hat {a} \, )  =
{\frac{1}{4}}\, \Gamma \, {\rm e}^{ - \Gamma \left| \deltat \right| }
\, \left[  \, 1 \, \pm \, {\cal {D}} \times \cos{ \deltamd \, \deltat } \,  \right]
\otimes {\cal {R}}( \, \deltat \, ; \, \hat {a} \, ) \
\end{equation}
where \dilution\ is the dilution factor (section~\ref{sec:tagging}) 
and ${\cal {R}}$ is the \deltat\ resolution (section~\ref{sec:deltat}) 

An unbinned maximum likelihood fit that takes into account individual event 
\deltat\ errors and tagging category is performed on 
events from the exclusively reconstructed \Bz\ sample (section~\ref{sec:exclusive}), 
after tagging (section ~\ref{sec:tagging}) has been performed.
The value of \deltamd\ is fitted simultaneously with the individual 
dilution factors for each tagging category. This information is later used in the 
\stwob\ extraction. 
Our preliminary result for \deltamd\ is : 
\begin{eqnarray*}
\deltamd =  0.516 \pm 0.031 ({\rm stat.}) \pm 0.018 ({\rm syst.}) \hbar \ps^{-1}
\end{eqnarray*}
A sample of events where a semileptonic ($D^*\ell\nu$) instead of a hadronic \Bz\ decay 
has been reconstructed (7517 events) are analysed using the same method and fit. 
The preliminary result for \deltamd\ from this sample is :
\begin{eqnarray*}
\deltamd = 0.508 \pm 0.020 ({\rm stat.}) \pm 0.022 ({\rm syst.}) \hbar \ps^{-1}
\end{eqnarray*}
Combining the \deltamd\ results from the hadronic and semileptonic
$B$ samples we obtain the preliminary result : 
\begin{eqnarray*}
\deltamd = 0.512 \pm 0.017 ({\rm stat.}) \pm 0.022 ({\rm syst.}) \hbar \ps^{-1}
\end{eqnarray*}
The main sources of systematic errors are the \deltat\ resolution function, 
Monte Carlo statistics and the \Bpm\ background in the semileptonic sample. 

In an independent analysis a more abundant but less pure sample of dilepton events has been used. 
In this inclusive approach the mistag arising from cascade leptons and the \Bpm\ fraction are 
extracted from the same fit as \deltamd. 
Our preliminary result for \deltamd\ is :
\begin{eqnarray*}
\deltamd  = 0.507 \pm 0.015 ({\rm stat.}) \pm 0.022 ({\rm syst.}) \hbar \ps ^{-1}  
\end{eqnarray*}
The results from the hadronic and dilepton samples are shown in Figure~\ref{fig:dm}.
\begin{figure}[htbp]
\vskip -.5cm
\begin{center}
 \mbox{\epsfig{file=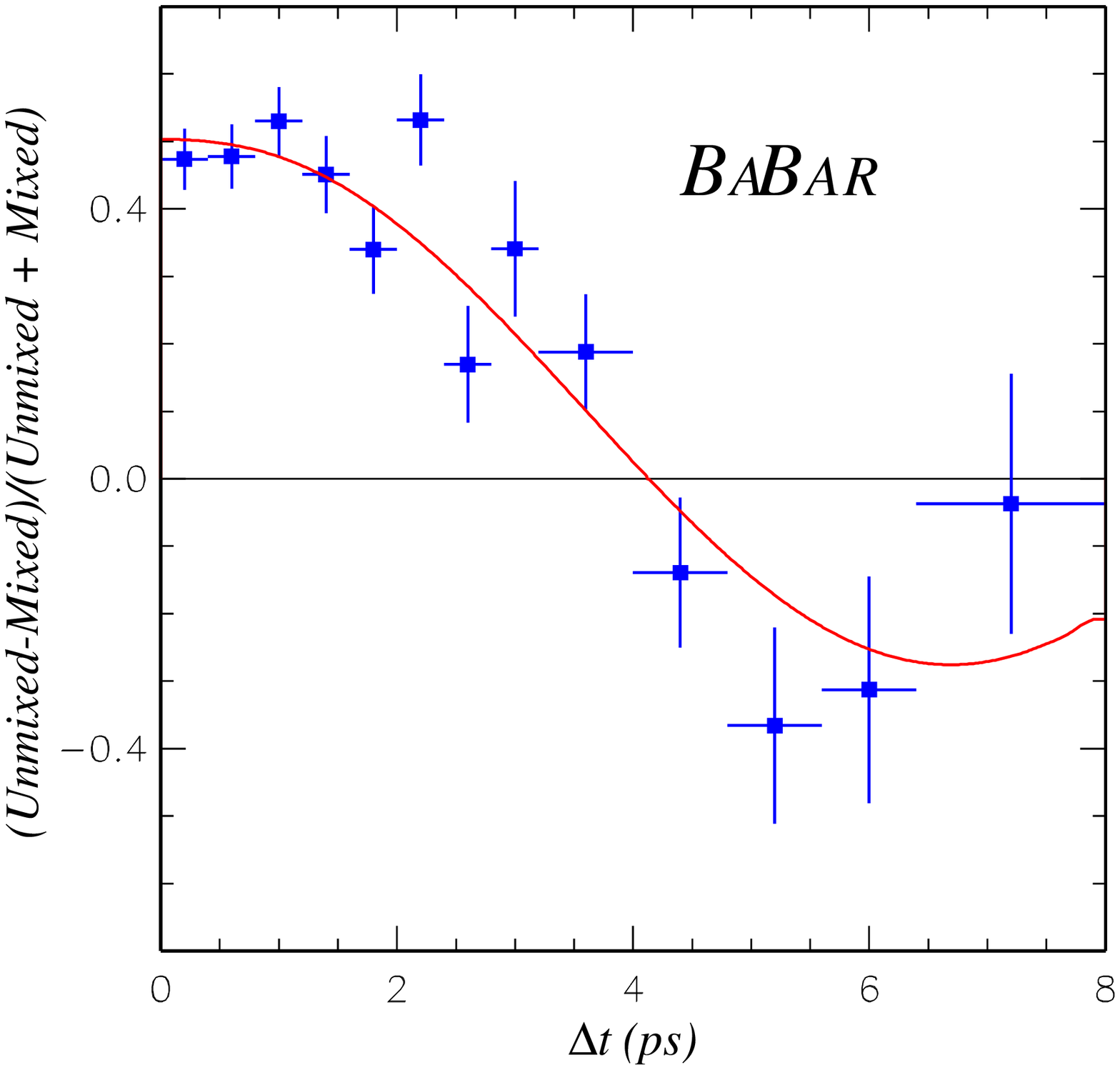,width=7.0cm}}
 \mbox{\epsfig{file=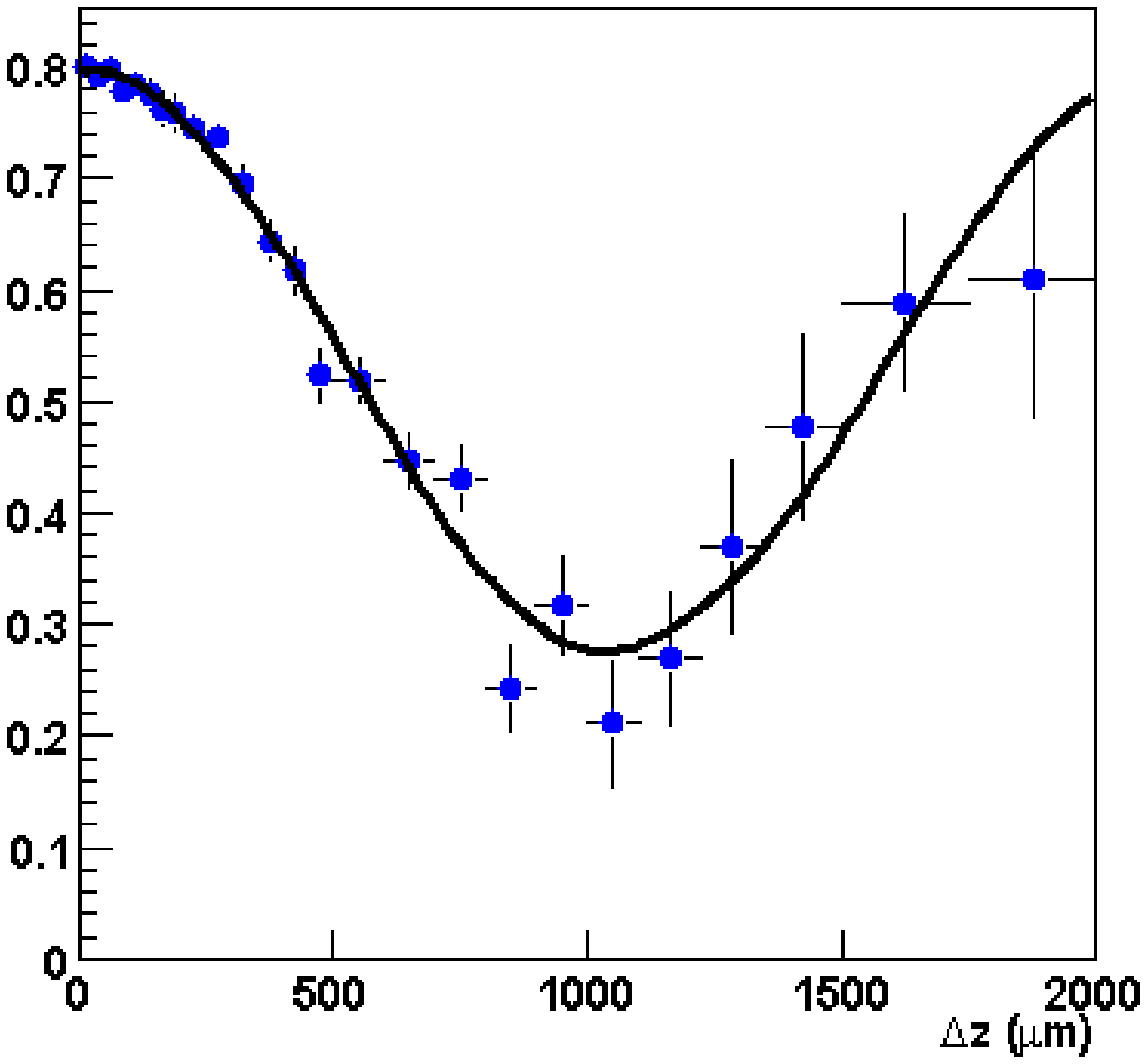,width=7.0cm}}
\end{center}
\caption{The observed time dependent asymmetries between unmixed and mixed events for the 
fully reconstructed (left) and dilepton (right) \Bz\ samples described in the text. 
The curves show the fit results.}
\label{fig:dm}
\end{figure}
The tagging performance parameters for each tagging method (category) 
are extracted from the fully reconstructed 
sample fits (hadronic and semileptonic) and are shown in Table~\ref{tab:mistag}.
\begin{table}[hb]
\vspace{0.3cm}
\begin{center}
\begin{tabular}{|l|c|c|c|} \hline
Tagging Category & $\varepsilon$ (\%) & $\mistag$ (\%) & $Q$ (\%)       \\ \hline \hline
{\tt Lepton}     & $11.2\pm0.5$ & $9.6\pm1.7\pm1.3$   &  $7.3\pm0.7$  \\
{\tt Kaon}       & $36.7\pm0.9$ & $19.7\pm1.3\pm1.1$  &  $13.5\pm1.2$  \\
{\tt NT1}        & $11.7\pm0.5$ & $16.7\pm2.2\pm2.0$  &  $5.2\pm0.7$  \\
{\tt NT2}        & $16.6\pm0.6$ & $33.1\pm2.1\pm2.1$  &  $1.9\pm0.5$  \\  \hline \hline
all              & $76.7\pm0.5$ &                     &  $27.9\pm1.6$ \\ 
\hline
\end{tabular}
\end{center}
\caption{Tagging preformance parameters measured from the mixing maximum-likelihood 
fit for the fully-reconstructed \Bz\ sample. The uncertainties on $\varepsilon$ and 
$Q$ are statistical only.}
\label{tab:mistag}
\end{table}

%%%%%%%%%%%%%%%%%%%%%%%%%%%%%%%%%%%%%%%%%%%%%%%%%%%%%%%%%%%%%%%%%%%
\section{The \stwob\ measurement}
If one of the neutral \B\ mesons( $B_{tag}$) of the coherent \BzBzb\ pair decays to a definite 
flavor eigenstate at time $t_{tag}$ and the other \B\ decays to a  \CP--even eigenstate 
like $\jpsi \KS$ or $\psitwos \KS$ at time $t_{CP}$, then in a perfect detector the 
following decay rates would be observed :
\begin{equation}
\label{eq:TimeDep0}
        f_\pm(\, \deltat \, ; \,  \Gamma, \, \deltamd, \, \sin{ 2 \beta } )  = 
{\frac{1}{4}}\, \Gamma \, {\rm e}^{ - \Gamma \left| \deltat \right| }
\, \left[ \, 1 \, \pm \, \sin{ 2 \beta } \times \sin{ \deltamd \, \deltat } \,  \right]
\end{equation}
where $\deltat = t_{CP} - t_{tag}$ and the ($+$) or ($-$) sign indicates whether the 
$B_{tag}$ is tagged as a \Bz\ or a \Bzb respectively. 
In the presense of the dilution factor \dilution\ 
and \deltat\ resolution ${\cal {R}}$ 
the observed rates become :
\begin{equation}
\label{eq:Convol}
{\cal F}_\pm(\, \deltat \, ; \, \Gamma, \, \deltamd, \, 
              {\cal {D}} \sin{ 2 \beta }, \, \hat {a} \, ) = 
{\frac{1}{4}}\, \Gamma \, {\rm e}^{ - \Gamma \left| \deltat \right| } \, 
\left[ \, 
1 \, \pm \, {\cal {D}} \sin{ 2 \beta } \times \sin{ \deltamd \, \deltat } 
\,  \right]
\otimes {\cal {R}}( \, \deltat \, ; \, \hat {a} \, )
\end{equation}
The time dependent decay rate asymmetry ${\cal A}_{CP}(\deltat)$ 
is a \CP--violating observable which (neglecting resolution effects) is 
approximately proportional to  \stwob:
\begin{equation}
{\cal A}_{CP}(\deltat) 
\, \, = \, \,  
\frac{ {\cal F}_+(\deltat) \, - \, {\cal F}_-(\deltat) }
{ {\cal F}_+(\deltat) \, + \, {\cal F}_-(\deltat) }
\, \, \sim \, \, 
{\cal D} \sin{ 2 \beta } \times \sin{ \deltamd \, \deltat }
\label{eq:asymmetry}
\end{equation}

%%%%%%%%%%%%%%%%%%%%%%%%%%%%%%%%%%%%%%%%%%%%%%%%%%%%%%%%%%%%%
\subsection{Analysis procedure}
The extraction of \stwob\ from the data follows the following steps~:
\begin{itemize}
\item
Selection of the signal $\Bz/\Bzb \to \jpsi \KS$ 
and $\Bz/\Bzb \to \psitwos \KS$ events, detailed in the following section. 
Backgrounds and in particular any admixture with the ``wrong'' \CP\ content 
have to be kept at a minimum level.
\item
Measurement of \deltat. The resolution is studied using simulated events and 
its parameters are actually extracted from real data, 
as described in section~\ref{sec:deltat}.
\item
Determination of the flavor of the $B_{tag}$, as described in 
section~\ref{sec:tagging}. 
The dilution factors ${\cal D}_i$ for each tagging category are measured on real 
data, as described in section~\ref{sec:mixing}.
\item
Extraction of the amplitude of the \CP\ asymmetry and the value of \stwob\ 
with an unbinned maximum likelihood fit, desribed in the following.
\end{itemize}
A blind analysis has been adopted for the extraction of 
\stwob.
A technique that hides the result of the fit by arbitrarily flipping 
its sign and adding an arbitrary offset, without affecting the error on the fitted 
parameters or their correlations, was used. 
Moreover, the visual \CP\ asymmetry in the  \deltat\ distribution is  
hidden by multiplying \deltat\ by the sign of the tag and adding an arbitrary offset.
Such an approach allows to optimise and finalise the event selection and 
fitting strategy as well as perform a variety of validation and stability 
checks without the posibility of any experimenter's bias.

%%%%%%%%%%%%%%%%%%%%%%%%%%%%%%%%%%%%%%%%%%%%%%%%%%%%%%%%%%%%%%%
\subsection{Event samples}
The \CP\ sample contains \Bz\ candidates reconstructed in the \CP\ eigenstates
$\jpsi \KS$ or $\psitwos \KS$. 
The charmonium mesons are reconstructed through their 
decays to \epem\ and \mumu, while the \psitwos\ is also reconstructed through its decay 
to $\jpsi\pipi$. The \KS\ is reconstructed through its decays to \pipi\
and \ppz.

Utilisation of the exclusively reconstructed \B\ samples (section~\ref{sec:exclusive}) 
for the characterisation of the tagging and vertexing performance and quality 
has already been described. 
In addition 570 $B^+ \to \jpsi K^+$ candidates and 
237 $\Bz \to \jpsi (K^{* 0} \to K^+ \pi^-)$ candidates have been 
reconstructed and used extensively in validation analyses.

%%%%%%%%%%%%%%%%%%%%%%%%%%%%%%%%%%%%%%%%%%%%%%%%%%%%%%%%%%%%%%
\subsection{Selection of events in the \boldmath \CP\ sample}
Events are required to have at least four reconstructed charged tracks, 
a vertex within 0.5\cm\ of the average position of the 
interaction point in the transverse plane, 
total visible energy greater than 5\gev, and 
second-order normalized Fox-Wolfram moment\cite{fox}  
($R_2=H_2/H_0$) less than 0.5. 

For the $\jpsi$ or $\psitwos \to \epem$ ($\mu^+ \mu^-$) candidates,
at least one of the decay products is required to be positively identified 
as an electron (muon) in the EMC (IFR). 
Electrons outside the acceptance of the EMC are accepted if their DCH \dedx\ 
information is consistent with the electron hypothesis.
Looser particle identification criteria are applied on the second electron (muon) 
candidates. 
In the muon case, a minimum ionising signature in the EMC is required.

$\jpsi$ candidates are selected with an invariant mass greater than 
2.95(3.06)\gevcc\ for $\epem$ ($\mu^+ \mu^-$) and  smaller than 
3.14\gevcc\ in both cases. 
The $\psitwos$ candidates in leptonic modes must have a mass 
within 50\mevcc\ of the  $\psitwos$ mass.
The lower bound is relaxed to 250\mevcc\ for the $\epem$ mode.
For the $\psitwos \to \jpsi \pipi$ mode, mass-constrained \jpsi\ candidates 
are combined with pairs of oppositely charged tracks considered as pions, 
and  $\psitwos$ candidates with mass between 3.0\gevcc\ and 4.1\gevcc\ are retained.  
The mass difference between the \psitwos\ candidate and the \jpsi\ candidate 
is required to be within 15\mevcc\ of the known mass difference.  

\KS\ candidates reconstructed in the $\pi^+ \pi^-$ mode are required to have
an invariant mass, computed at the vertex of the two tracks, 
between 486\mevcc\ and 510\mevcc\ for the $\jpsi \KS$ selection, 
and  between 491\mevcc\ and 505\mevcc\ for the $\psitwos \KS$ selection.   

For the  $\jpsi \KS$ mode we also consider the decay of the \KS\ 
into $\pi^0 \pi^0$. 
For pairs of \piz\ candidates with total energy above 800\mev 
we determine the most probable \KS\ decay point along the path defined 
by the \KS\ momentum vector and the primary vertex of the event.
The decay-point probability is the product of the $\chi^2$ probabilities 
for each photon pair constrained to the $\pi^0$ mass.
We require the distance from the decay point to the primary vertex 
to be between $-10$~cm and $+40$~cm and 
the \KS\ mass measured at this point to be between 470 and 536\mevcc.

$B_{CP}$ candidates are formed by combining mass-constrained \jpsi\ or
\psitwos\ candidates with mass-constrained \KS\ candidates.
Cuts on the colinearity of flight vertex and momentum of the \KS\ (for 
\pipi\ decays), the cosine of the helicity angle of the \jpsi\ or \psitwos\
in the \B\ candidate rest frame (\epem\ and $\mu^+ \mu^-$ modes) or 
the  cosine of the angle between the $B_{CP}$ candidate three-momentum 
vector and the thrust vector of the rest of the event 
($\psitwos \to \jpsi \pipi$ mode) are applied to achieve the required signal 
purity. 

$B_{CP}$ candidates are identified in the \mes--${\rm \Delta} E$ plane 
(see section~\ref{sec:exclusive}). 
Signal event yields and purities, determined from a fit to the \mes\ 
distributions after selection on ${\rm \Delta} E$, are presented in
Table~\ref{tab:CharmoniumYield}. The \CP candidate events are 168 with 
a purity of 95.6\%. In 120 of these events there is information on the 
flavor of the other \B. These events are used in the final fit for \stwob.
\begin{table}[htbp]
\begin{center}
\begin{tabular}{|l|l|c|c|} \hline
Final state  & All events & Purity  & Tagged events\\ \hline \hline
$\jpsi \KS$ ($\KS \to \pi^+\pi^-$)   &  124 & 96\% & 85 \\
$\jpsi \KS$ ($\KS \to \pi^0 \pi^0$)  &   18 & 91\% & 12 \\
$\psitwos \KS$                       &   27 & 93\% & 23 \\ 
\hline \hline
\end{tabular}
\end{center}
\caption{ Event yields for the \CP\ samples used in this analysis. 
The total number of events in the \mes--${\rm \Delta} E$ signal box 
and their purity, as well as the size of the subsamples where the 
other \B\ is tagged, are shown.}
\label{tab:CharmoniumYield}
\end{table}
Distributions of ${\rm \Delta} E$ and \mes\ are shown in 
Figures~\ref{fig:jks} and~\ref{fig:psi2sks}.
\begin{figure}[htbp]
\begin{center}
 \mbox{\epsfig{file=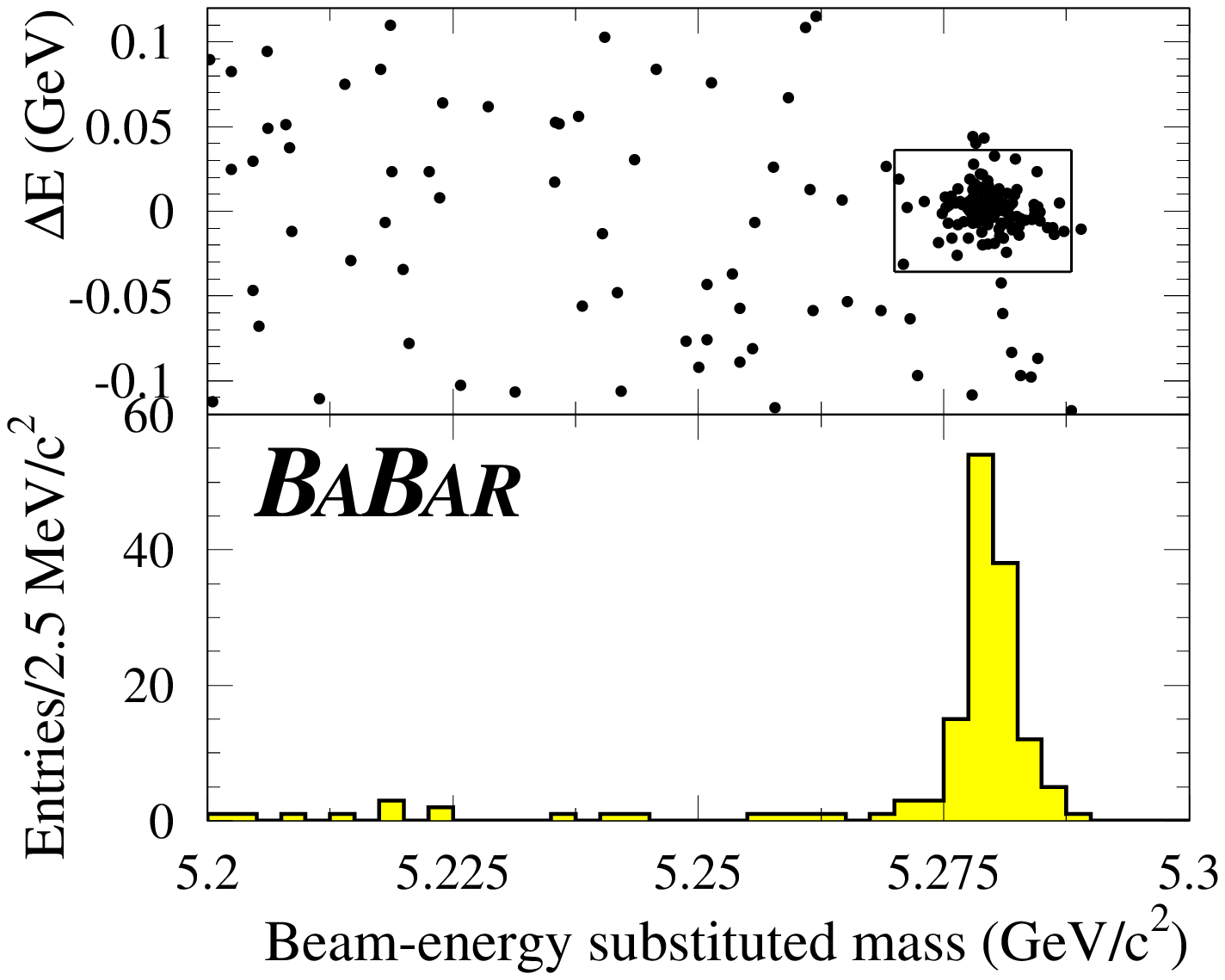,height=7.5cm}}
 \mbox{\epsfig{file=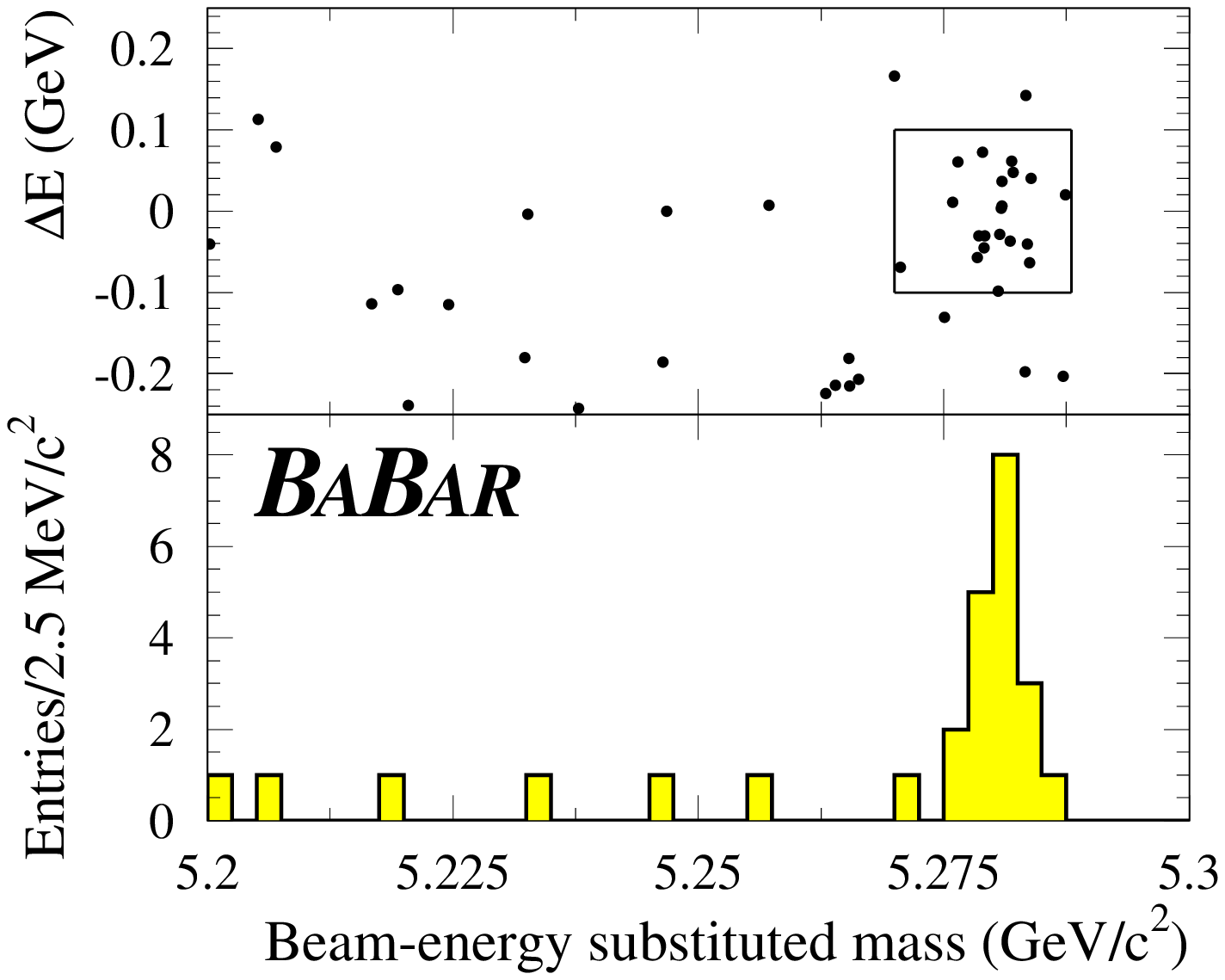,height=7.5cm}}
\caption{$\jpsi\KS$ signal. Left: $\KS \to \pipi$, Right: $\KS \to \ppz$ }
\label{fig:jks}
 \includegraphics[height=7.5cm]{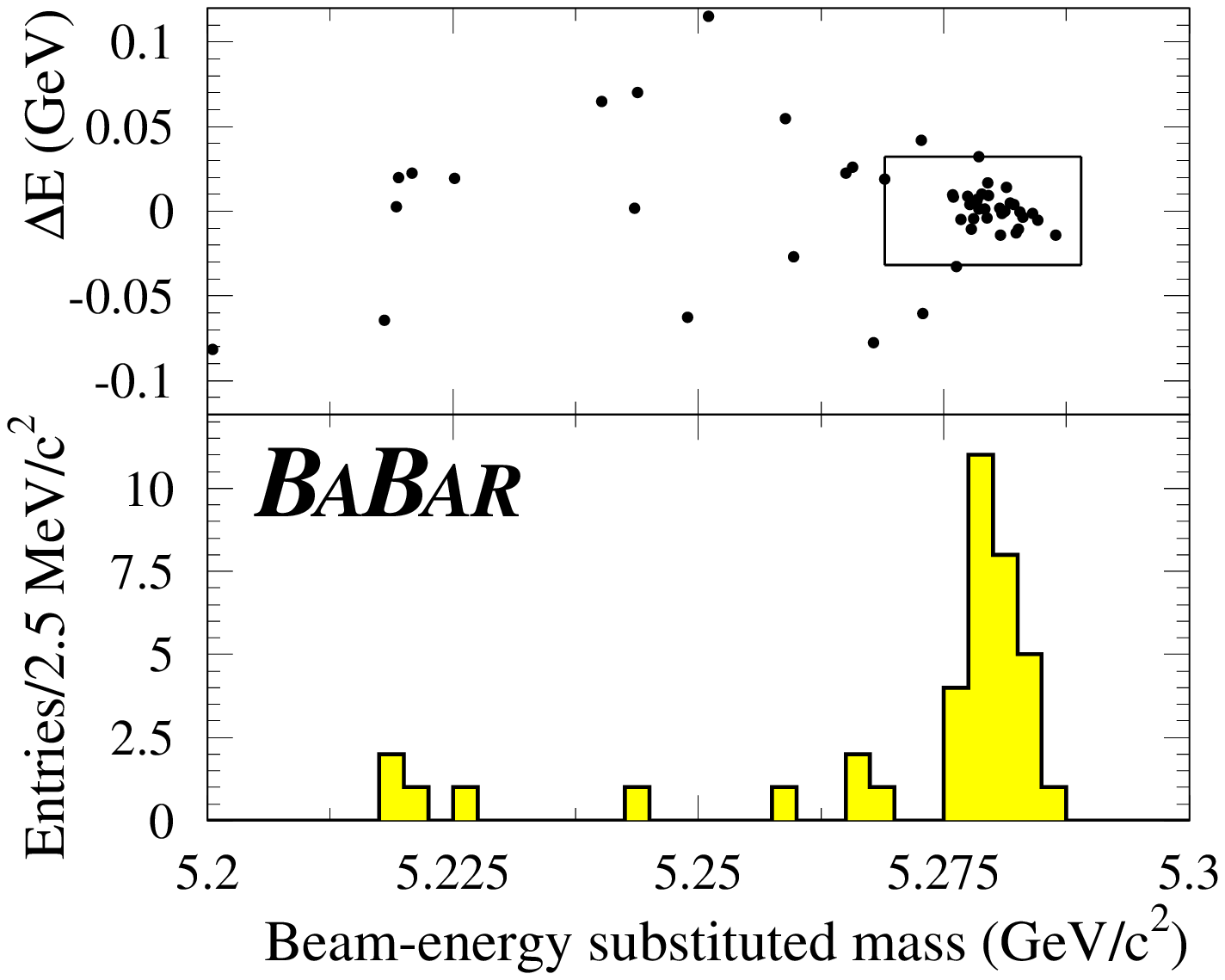}
\caption{$\psitwos \KS$ ($\KS \to \pi^+ \pi^-$)  signal.}
\label{fig:psi2sks}
\end{center}
\end{figure}

%%%%%%%%%%%%%%%%%%%%%%%%%%%%%%%%%%%%%%%%%%%%%%%%%%%%%%%%%%%%%%%%
\subsection{Extracting \stwob}
The \deltat\ of the 120 selected and tagged events is fitted to the 
expected time evolution (equation~\ref{eq:Convol}) with an unbinned 
maximum likelihood fit. Individual event errors on \deltat\ are taken into 
account. The resolution determined on the fully reconstructed sample and the 
mistag factor \mistag$_i$ corresponding to the tagging category for each event 
are used in the fit. The \deltamd\ are fixed to the nominal PDG~\cite{bib:PDG2000} 
values of $\tau_{\Bz} = 1.548\ \ps$ and $\deltamd = 0.472 \, \hbar \ps^{-1}$ 
respectively. The resulting errors on \stwob\ due to these uncertainties 
are 0.002 and 0.015.

%%%%%%%%%%%%%%%%%%%%%%%%%%%%%%%%%%%%%%%%%%%%%%%%%%%%%%%%%%%%%%%%%%
\subsection{Fit validation, systematics studies and null \CP\ tests}
Knowledge of the mistag fractions, description of the \deltat\ resolution and 
backgrounds are (in that order) the main sources of systematic errors. All these 
have been extracted from real data. Real data, fully simulated 
Monte Carlo, and ``Toy'' Monte Carlo samples have been used to validate the method 
and implementation of the fit, to rule out possible biases from the method itself, 
and to assess the size of systematic errors.

The full CP analysis and fit were performed on data samples that 
have no \CP\ asymmetry. No significant apparent \CP\ effect was measured, as 
shown in Table~\ref{tab:nulltest}. The 1.9 $\sigma$ asymmetry 
in the $\jpsi K^{*0}$ channel is interpreted as a statistical fluctuation. 
\begin{table}[htbp]
\begin{center}
\begin{tabular}{|l|c|} \hline
 Sample   & Apparent \CP-asymmetry  \\ \hline \hline
 Hadronic charged \B\ decays &   $0.03 \pm 0.07$   \\  \hline
 Hadronic neutral \B\ decays &   $-0.01 \pm 0.08$  \\  \hline
 $\jpsi K^+$                 &   $0.13\pm 0.14$    \\ \hline 
 $\jpsi K^{*0}$ ($K^{*0} \to K^+ \pi^-$) &  $ 0.49 \pm 0.26$ \\ \hline 
\end{tabular}
\end{center}
\caption{Results of fitting for apparent \CP\ asymmetries in various 
charged or neutral flavor-eigenstate \B\ samples. } 
\label{tab:nulltest}
\end{table}

\clearpage

%%%%%%%%%%%%%%%%%%%%%%%%%%%%%%%%%%%%%%%%%%%%%%%%%%%%%%%%%%%%%%%%%%
\subsection{Results}
The maximum-likelihood fit for \stwob\ on the full tagged sample of 
$\Bz/\Bzb \to \jpsi\KS$ and $\Bz/\Bzb \to \psitwos \KS$ events 
yields the preliminary result :
\begin{equation}
\result
\end{equation}
The results of the fit for each type of \CP\ sample and for each 
tagging category are given in Table~\ref{tab:result}.  
The contributions to the systematic uncertainty are summarized
in Table~\ref{tab:systematics}.
The \deltat\ distributions for \Bz\ and \Bzb\ tags are shown 
in Fig.~\ref{fig:deltatfit} and the raw asymmetry as 
a function of \deltat\ is shown in Fig.~\ref{fig:asymmetry}.
The probability of obtaining a value of the statistical error 
larger than the one we observe is estimated at 5\%.
Based on  a large number of full Monte Carlo simulated experiments with the 
same number of events as our data sample, we estimate that the probability 
of finding a lower value of the likelihood than our observed value is 20\%. 
\begin{table}[htbp]
\begin{center}
\begin{tabular}{|l|c|} \hline
sample                                    &  \stwob  \\ \hline \hline
\CP\ sample                               &  {\bf 0.12}$\pm${\bf 0.37}  \\  
\hline
\ \ $\jpsi \KS$ ($\KS \to \pi^+ \pi^-$) events  &  $-0.10 \pm 0.42$   \\  
\ \ other \CP\ events                           &  $0.87 \pm 0.81$   \\  
\hline
\ \ {\tt Lepton}                         &  $1.6 \pm 1.0  $   \\
\ \ {\tt Kaon}                           &  $0.14\pm 0.47   $ \\
\ \ {\tt NT1}                            &  $-0.59\pm0.87  $  \\
\ \ {\tt NT2}                            &  $-0.96\pm 1.30  $ \\
\hline 
\end{tabular}
\end{center}
\caption{\stwob\ fit results from the entire \CP\ sample and 
various subsamples.} 
\label{tab:result}
\end{table}
\begin{table}[htbp]
\begin{center}
\begin{tabular}{|l|c|} \hline
Source of uncertainty    &  Uncertainty on \stwob \\ \hline \hline 
uncertainty on $\tau_\Bz$                           &   0.002     \\
uncertainty on \deltamd                             &   0.015     \\
uncertainty on \deltaz\ resolution for \CP\ sample  &   0.019     \\ 
uncertainty on time-resolution bias for \CP\ sample &   0.047     \\ 
uncertainty on measurement of mistag fractions      &   0.053     \\   
different mistag fractions for \CP\ and non-\CP\ samples &   0.050 \\
different mistag fractions for \Bz\ and \Bzb\       &   0.005     \\
background in \CP\ sample                           &   0.015     \\
\hline \hline
total systematic error                            & {\bf 0.091 }  \\ 
\hline 
\end{tabular}
\end{center}
\caption{ Summary of systematic uncertainties on \stwob}
\label{tab:systematics}
\end{table}
\begin{figure}[htbp]
\begin{center}
\includegraphics[height=8cm]{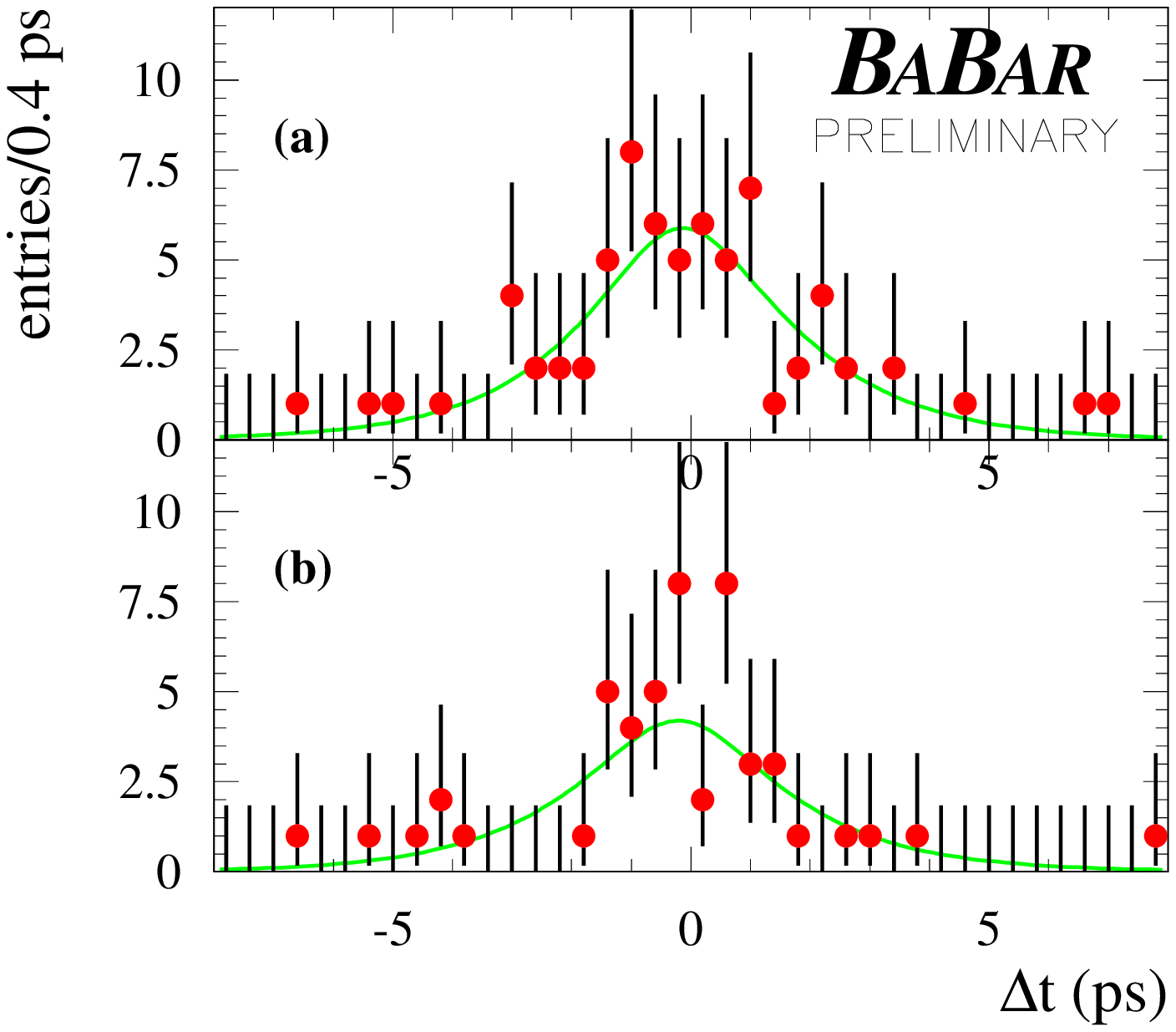}
\vspace{-1cm}
\caption{ Distribution of \deltat\ for (a) the \Bz\ tagged events and 
(b) the  \Bzb\ tagged events in the \CP\ sample. 
The error bars plotted for each data point assume Poisson statistics.  
The curves correspond to the result of the unbinned maximum-likelihood fit
and are each normalized to the observed number of tagged \Bz\ or \Bzb\ events.}
\label{fig:deltatfit}
\end{center}
%\end{figure}
%%
%\begin{figure}[htbp]
\begin{center}
\includegraphics[height=7cm]{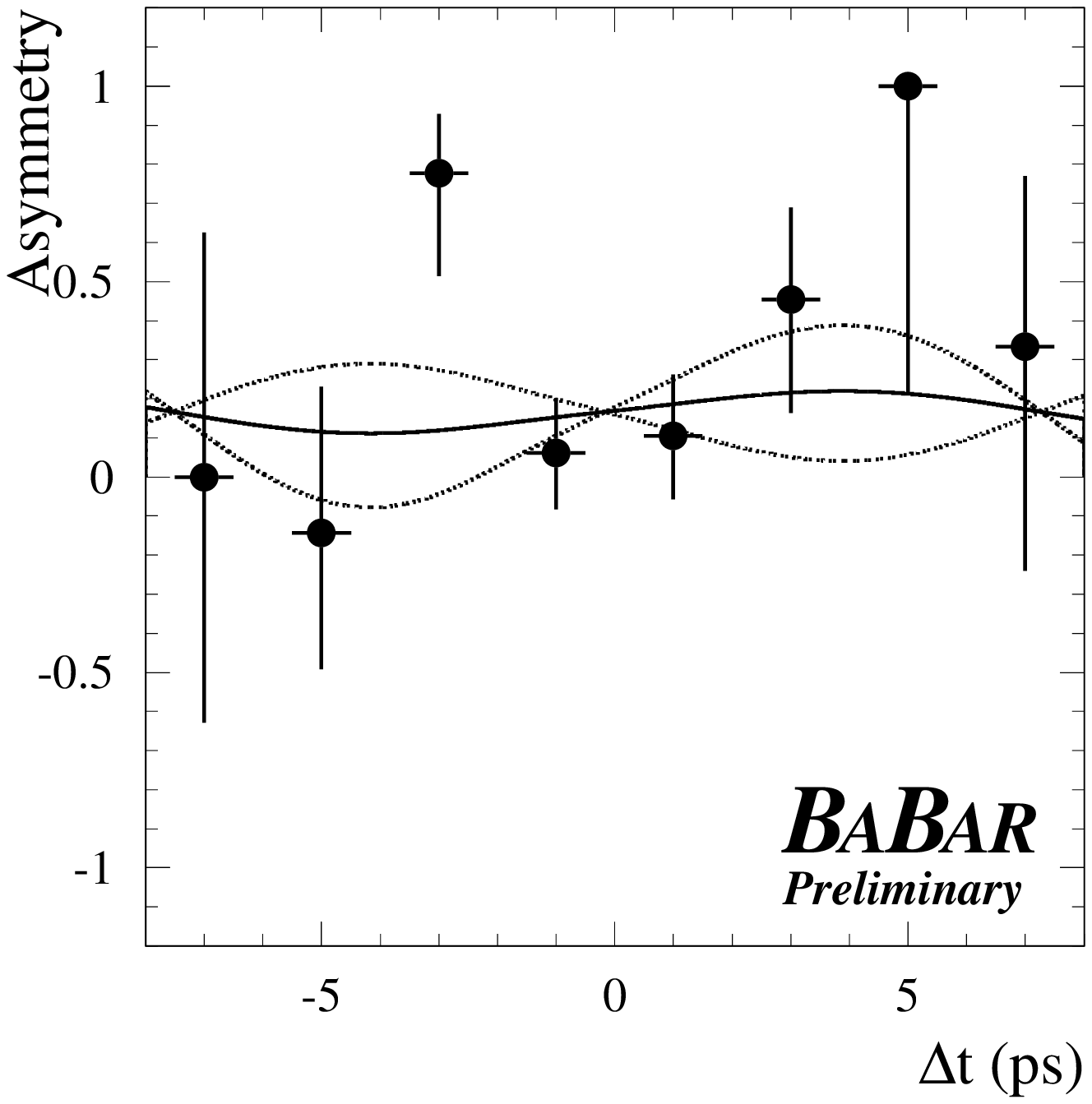}
\caption{The raw \Bz-\Bzb\ 
asymmetry $(N_{\Bz}-N_{\Bzb})/(N_{\Bz}+N_{\Bzb})$ with binomial errors 
as function of \deltat.  
The solid curve represents our central value of \stwob. 
The two dotted curves correspond to 
one statistical standard deviation from the central value.
The curves are not centered at $(0,0)$ in part because 
the probability density functions are normalized 
separately for \Bz\ and \Bzb\ events, and
our \CP\ sample contains an unequal number of 
\Bz\ and \Bzb\ tagged events (70 \Bz\ versus 50 \Bzb).
The $\chi^2$ between the binned asymmetry and the result of the 
maximum-likelihood fit is 9.2 for 7 degrees of freedom. }
\label{fig:asymmetry}
\end{center}
\end{figure}

\clearpage

%%%%%%%%%%%%%%%%%%%%%%%%%%%%%%%%%%%%%%%%%%%%%%%%%%%%%%%%%%%%%%%
\section{Conclusions and prospects}
The first \babar\ measurement of the \CP-violating asymmetry parameter \stwob 
has been presented~:
\begin{equation}
\result \, \, \, \, (preliminary)
\end{equation}  
\babar\ aims at collecting more than 20\invfb of data by the end of Run 1 in fall 
2000. A measurement of \stwob with a precision better than 0.2 is expected early  
in 2001.

Very competitive preliminary results have also been presented for the 
\B meson lifetimes, as well as the first measurements of \BzBzb mixing at the \FourS. 
These measurements will also benefit in the near future from the expected 
significant increase in statistics.

%%%%%%%%%%%%%%%%%%%%%%%%%%%%%%%%%%%%%%%%%%%%%%%%%%%%%%%%%%%%%%


\begin{thebibliography}{99}

\bibitem{Jarlskog}
C.\ Jarlskog, in {\em CP Violation}, C. Jarlskog ed., World
Scientific, Singapore (1988).

\bibitem{BabarPub0018}
\babar\ Collaboration, B.\ Aubert {\em et al.},
``The first year of the \babar\ experiment at \pep2'',
\babar-CONF-00/17, submitted to the XXX$^{th}$
International Conference on High Energy Physics, Osaka, Japan.

\bibitem{bib:argusfunction} ARGUS collaboration, H.\ Albrecht {\em et al.}, 
Z~Phys.~{\bf C48}, 543 (1990).

\bibitem{fox}
G.~C.~Fox and S.~Wolfram, \jprl{41}, 1581 (1978).

\bibitem{bib:PDG2000} 
Particle Data Group, D.\ E.\ Groom {\em et al.},
\epjc{15}, 1 (2000).

\end{thebibliography}
\end{document}